# Digital Fragmentation and Generative AI Use Across 103 Million Application Events

**Authors:** Sumer S. Vaid[1*] and Ashley V. Whillans[2*]

**Affiliations:**

[1]Artificial Intelligence Institute, Harvard Business School

[2]Negotiation, Organizations and Markets Unit, Harvard Business School

*Corresponding author. Email: svaid@hbs.edu; awhillans@hbs.edu

**Abstract:** Knowledge workers switch between applications thousands of times per day, spending nearly a tenth of the work year transitioning between digital applications in a process called digital fragmentation. Whether this fragmentation reflects who an employee is, where they work, or what kind of day they are having, has remained an open question. We analyzed 103 million application events recorded second-by-second from 1,017 employees across eight organizations that largely employ knowledge workers (e.g., law, financial services). Day-to-day variation in fragmentation within individual employees accounted for 44.6% of the variation in digital fragmentation, slightly exceeding stable individual differences between employees (35.8%), and far exceeding variation between organizations (19.6%). Fragmentation rose over the work week and reset after weekends and holidays. Higher-than-typical use of communication applications coincided with more fragmented work. Generative AI use also occurred on more fragmented days, but the period following AI use was marked by narrower, longer, and more predictable application use. These findings identify the workday as a key level for understanding and intervening on digital fragmentation and suggest that AI may help structure fragmented work rather than merely intensify it.

Knowledge work, the application and exchange of information[1], primarily unfolds inside digital applications such as email, spreadsheets and, increasingly, AI tools. Knowledge workers move between these applications and tools in rapid succession, and they rarely remain within a single application for more than a few minutes at a time. We refer to these rapid switches between applications[2] as "digital fragmentation." By some estimates, knowledge workers toggle between applications up to 1,200 times each day, spending nearly 9% of their work year in transition[3].

Digital fragmentation requires employees to reorient after each switch and is associated with poor work outcomes. A high frequency of application switching is associated with long task completion times[4], high error rates[5], and increased difficulty resuming interrupted work[6–10]. Moreover, digital fragmentation has intensified over time. According to research that has tracked workers' activities using direct in-person observation and activity logs, the amount of uninterrupted time that employees spend on individual applications has dropped from about 2.5 minutes two decades ago[2] to 47 seconds in the last decade[11].

The time it takes for employees to reorient between applications[9] accumulates to roughly 5 weeks of productive output per year per employee[3], a magnitude that, multiplied across the global knowledge workforce, translates into hundreds of billions of dollars in lost productive activity each year.

Despite mounting evidence that digital fragmentation is harmful and widespread[2–9], a fine-grained and representative account of digital fragmentation has remained elusive. The largest studies of the digital workday have relied on aggregated measures of digital fragmentation such as email volume and meeting counts over days or weeks[12–14], which cannot show how employees move between applications on a momentary, second-by-second basis. Other studies that have examined switching at a more granular time scale confirm that it is pervasive[15–19]. However, these studies rely on small samples that cannot capture variation across people or organizations. The observation windows deployed in these studies typically run from about five days[15] to two weeks, which are too short to study patterns of switching that occur over weeks and months[20]. These limitations leave three central questions about digital fragmentation unanswered.

First, where does digital fragmentation come from? At any given time, application switching could arise from at least three distinct sources – differences between employees[7,10] in how they work, day-to-day variation in an employee's work demands[21], and organizational and industrial norms[21]. Because prior research has only tracked a small number of employees within a single organization over a limited time window[15–19], existing evidence cannot adjudicate between these distinct sources of fragmentation. Indeed, understanding where digital fragmentation comes from requires studying many employees across many days and organizations.

Second, how does digital fragmentation change over time? Studies have rarely tracked the same employees across multiple continuous weeks (to the best of our knowledge, the longest study tracked 20 software developers for an average of 11 consecutive workdays[19]). Thus, it is unclear how digital fragmentation shifts across the workweek, whether weekends and holidays reset day-to-day patterns of fragmentation, and whether digital fragmentation persists across days.

Third, how does generative AI use coincide with digital fragmentation? Nearly all of the available data on digital fragmentation was collected before generative AI tools entered the workplace[22]. To date, the studies that draw on post-AI data have examined the effect of AI on worker productivity and output quality rather than examining how the use of AI coincides with application switching[23–25]. Consequently, we do not know whether the findings on digital fragmentation that were documented in earlier studies continue to describe digital fragmentation as it occurs today.

To address these questions, we draw on a dataset of 103 million application events recorded from 1,017 knowledge workers across eight organizations (spanning software development, accounting, financial services, e-publishing, law, and marketing), who were recorded second-by-second over a median of 58 workdays. The scale, granularity, and breadth of these data bring digital fragmentation into view at the timescale in which it occurs. These data allow us to examine how digital fragmentation differs from one *employee* to the next, allowing us to assess whether some employees consistently switch between applications more than others. These data also allow us to understand the role of the *workday*, allowing us to ascertain how digital fragmentation changes from one day to the next within the same employee, and whether it follows recognizable rhythms across the workweek. Because we collect data across eight organizations that vary in their business goals and outputs (e.g., finance, consulting, and technology), we are also able to examine differences in fragmentation across knowledge-work organizations.

**Preview of the Results**

We report three sets of findings that correspond to each of the questions raised above.

First, we show that digital fragmentation differs substantially between employees, but it differs even more from one day to the next *within* the same employee. Stable differences across employees account for 35.8% of the variation in digital fragmentation we observe, while day-to-day differences within the same employee account for 44.6% of the variation, and only 19.6 % of this variation is explained by differences *between* organizations.

Second, we show that day-to-day changes in digital fragmentation within employees follow a predictable pattern; digital fragmentation accumulates across consecutive workdays, peaks during midweek, and partially resets across weekends and national holidays.

Third, we show that the use of communication tools as well as the use of generative AI tools coincide with day-to-day changes in digital fragmentation. On days with moderately elevated communication, digital fragmentation rises above an employee's typical level; on days when communication comes to dominate the workday, it falls. On more fragmented days, employees tend to use generative AI more often. However, in the window immediately after using an AI tool, employees' application use becomes more consolidated and predictable: they move through fewer unique applications, switch less often, spend longer in each application, and concentrate their activity more heavily in their most-used applications.

**What does digital fragmentation look like?**

Before presenting the results of our primary analyses, we turn our attention to characterizing digital fragmentation as it appears in our data across employees, workdays, and organizations. Throughout the manuscript, a workday denotes a weekday (Monday to Friday) on which an employee logged application activity, with public holidays additionally excluded where stated; all recorded activity from 00:00 to 23:59 contributes to each day's measures (Supplementary Table S1).

At the level of the individual employee, digital fragmentation manifests at a second-by-second resolution across the workday. We note that it is pervasive across employees and organizations (Fig. 1). Across our sample of 103 million application events, employees made 56.9 million switches over 62,693 person-days of observation, moving between applications a mean of 908.3 times per day (00:00 to 23:59) and rarely had a single application in the foreground of their computer screens for more than a few seconds. Instead, employees' digital media use interleaves productivity, communication, and AI tools.

Employees varied considerably in how often they switched between applications. The median employee made 814 application switches per workday; the mean employee made 943, and the middle 50% of employees ranged from 547 to 1,130 switches per workday. The most fragmented employees switched 4.5 times as often as the least fragmented, so that some employees operated in a near-continuously fragmented state, while others sustained long stretches within a single application. This variation extended within employees: a typical employee's daily switching moved by roughly 383 switches around their own average from one workday to the next, such that the same employee appeared heavily fragmented on one day and consolidated on the next. Digital fragmentation therefore varied both across employees and, within each employee, from day to day.

Our sample consists of eight organizations spanning six industries: two software firms (a software-development firm, $N = 49$; a software-testing firm, $N = 22$), two accounting firms ($N = 124$ and $N = 197$), one financial-services firm ($N = 212$), one e-publishing firm ($N = 237$), one law firm ($N = 99$), and one marketing firm ($N = 38$). Time spent across different applications varied substantially between organizations. Organizations differed in which applications their employees used and for how long, yet fragmentation itself was distributed similarly across them: differences between organizations accounted for 19.6% of the variance, with the remainder reflecting stable differences between employees (35.8%) and day-to-day variation within employees (44.6%).

Together, these descriptive patterns characterize the digital fragmentation observed across the sample. Most employees experience substantial fragmentation across the workday, moving between applications hundreds to more than a thousand times per day, though the degree of fragmentation varies across employees, across the hours of a single workday and, to a lesser extent, across organizations. Employees across the eight organizations draw on a similar set of digital applications and exhibit comparable patterns of fragmentation. The sections that follow examine this fragmentation formally, beginning with a measure that captures its structure.

**A Novel Measure of Fragmentation: The Fragmentation Index**

To better understand the digital fragmentation that we observe, we construct a novel measure that we call the Fragmentation Index. Previous characterizations of switching behaviors have largely relied on count data about the number of meetings and emails per day[13,14] and aggregated session duration data[26]. These coarse measures treat every application switch as equivalent, and discard information about which applications flow into one another. However, a switch from a code editor to a messaging application reflects a different kind of transition as compared to a switch between a code editor and Word processing software. We argue that treating these switches as interchangeable obscures the rich structure of the fragmented workday.

Building upon this past research, our Fragmentation Index captures the overall breadth of application switching across a workday (how many different applications an employee moves between in a given day), the unpredictability of application switching across a workday (how hard it is to anticipate which application comes next), and brevity (how short an employee's uninterrupted stretches in a single application are).

The Fragmentation Index analyzes digital application usage via a network approach[27] and captures how switching is organized, distinguishing between diffuse movement across many pairs of applications and clustered movement through frequently co-used pairs of applications. Separately, we capture fragmentation via a range of secondary measures that provide a granular view into the specific digital behaviors that differ across people and contexts, such as switching rate (how often an employee changes applications) and the number of productivity app interruptions (how often focus in a productivity application is punctuated by a switch to another application such as email or chat). See Table S1 for an explanation of each Fragmentation Index component.

First, as part of this Index, we calculate the unpredictability of application switching across a workday (how hard it is to anticipate what application employees will use next). This is a Shannon entropy measure[28] of how unpredictable an employee's application switches are across a workday. Across our sample, unpredictability was high overall and varied widely from one workday to the next (median 33.9, IQR 24.1 to 44.4), indicating that application switches were largely unpredictable. To our knowledge, transition entropy of application switching has not previously been estimated for desktop work; the closest prior work characterizes app-switch networks on smartphones[28] but does not report a comparable entropy value. We therefore benchmark each workday against the application pairs an employee used that day, asking how unpredictable the next switch was given the routes the employee relied on. On average, observed entropy reached 81% of its maximum possible value, meaning that among the trajectories of application taken by an employee, the next application was nearly as unpredictable as a random switch.

Second, as part of this Index, we calculate the breadth of application switching via network edge count[29]. This is the number of distinct switching pathways an employee uses across a workday. Network edge counts were large overall and varied widely between workdays (median 118, IQR 81 to 159), indicating that most workdays drew on a broad set of application pairings and that the number of pairings shifted substantially from one day to the next. The most common pathway connected File Explorer and Microsoft Excel, accounting for roughly 3.14 million transitions, and File Explorer appeared in seven of the ten most common pairings, marking it as a hub in employees' workflows. Switching nonetheless remained largely within productive work: 73.8% of transitions connected one productive application to another, and 2.6% led from a productive application to a non-productive application.

Third, as part of this Index, we calculate the mean length of an employee's continuous bouts of application use across a workday (a bout is a sequence of application uses bounded by gaps of at least five minutes), which we reverse-score so that shorter bouts indicate greater fragmentation. Bouts were long overall and varied widely from one workday to the next (median 80.5 minutes, IQR 50.8 to 130.0). However, switching was rapid within individual bouts: an application held an employee's focus for a mean of 27.8 seconds before the next switch, and the distribution was heavily right-skewed, with a median focus event of only 3.3 seconds and a small share of longer stretches accounting for most of the elapsed time. At 27.8 seconds, focus durations were roughly 40% shorter than the 47 seconds reported in the most recent study of digital work[11] and a fraction of the 2.5 minutes reported two decades ago[2].

These three indicators load onto a single factor that explains 66.8% of variance (PC1), which supports the treatment of the Fragmentation Index as a single dimension of digital fragmentation at the person-day level. A higher Fragmentation Index value therefore indicates a workday composed of more unpredictable switching across a broader set of application pairings, with shorter continuous bouts in any one application.

The Fragmentation Index varied across the hours of the workday with two distinct peaks, rising from an early morning trough (07:00 – 08:00) to a midmorning peak **(**10:00 to 12:00, highest at 11:00**)**, falling to a midday dip (13:00 – 14:00)**,** and rising again to a smaller late-afternoon peak before declining toward the end of the day (15:00 to 17:00, highest at 17:00**;** Fig. 4A).

At the person-day level, the Fragmentation Index was distributed unimodally around its sample mean, with most observations clustering near the center and a longer right tail capturing the more heavily fragmented end of the distribution (Fig. 4B). At the person-day level, most workdays clustered near the center of the distribution, with a smaller number of intensely fragmented days forming a longer tail (Fig. 4B). On the least fragmented days, employees moved through a handful of applications in long, uninterrupted stretches (fewer than 201 switches across roughly 15 or fewer applications); on the most fragmented, they switched constantly across many applications, rarely settling in any one for long (more than 1,870 switches across 38 or more applications).

We use the Fragmentation Index to address the first question raised in the introduction: whether digital fragmentation operates at the level of the workday, the employee, or the organization. We fit an unconditional random-intercept model that partitions the total variance in the Fragmentation Index across these three nested levels[30], and expressed each level's share as a proportion of the total (restricted maximum likelihood; see Methods). This decomposition follows the standard approach for intensive longitudinal data, in which repeated daily observations separate stable between-person differences from within-person variation across days[31]. Across 52,882 person-days from 978 employees, day-to-day differences within the same employee accounted for 44.6% of the variance, stable differences between employees for 35.8%, and differences between organizations for 19.6% (Fig. 4C). Furthermore, the Fragmentation Index was moderately stable from week to week within employees (ICC = 0.71). This result was robust under alternative data-construction choices, including the session-inclusion threshold and the data cleaning rules (Supplementary Tables S29 and S45)

**Digital fragmentation moves with the rhythms of the workweek**

Having established that digital fragmentation moves substantially from one day to the next, we turned to the second question: how is fragmentation organized across the workweek?

We fit multilevel models that regressed each day's fragmentation on the previous day's fragmentation within person, with random intercepts for employee and organization.

Digital fragmentation persisted over consecutive workdays. Across 48,319 consecutive-day pairs, an employee's Fragmentation Index on any given day predicted its level on the day that followed (within-person $\beta = 0.21$, $P < .001$). This result was robust after controlling for day of week ($\beta = 0.21$, $P < .001$; Supplementary Table S12).

The level of fragmentation fluctuated across the workweek, peaking on Tuesday before declining toward Friday (Fig. 5A). The weekend interrupted this momentum in the Fragmentation Index (Supplementary Table S13). We extended the carryover models to allow the day-to-day association in digital fragmentation to differ across each pair of adjacent days. The Friday-to-Monday autocorrelation was weaker than the Thursday-to-Friday autocorrelation for the Fragmentation Index (interaction $\beta = -0.086$, $P < .001$). The weekend therefore served as a partial reset of the digital fragmentation an employee accumulated across the workweek.

Indian national holidays during the study window (five of the 37 in our two-year reference calendar; Methods) coincided with a similar reset around longer breaks. We aligned each employee's Fragmentation Index to the workdays surrounding each holiday, excluding dates that fell within five workdays of two or more holidays so that each day is estimated from isolated holidays. Fragmentation dipped on the eve of each holiday (within-person $\beta = -0.13$, $P < .001$) and dropped further on the first workday afterward ($\beta = -0.46$, $P < .001$), returning to baseline within two workdays ($\beta = -0.02$, n.s., by day +2). The dip was specific to the days immediately adjacent to the holiday rather than a gradual decline across the surrounding week; estimates at the outermost window days rest on the thinned isolated-holiday boundary and are reported in full in Supplementary Table S16. The eve-of-holiday and first-day-back dips held after adjusting for day of week ($\beta = -0.12$ and $-0.40$, both $P < .001$; Supplementary Table S16).

Day-to-day fragmentation therefore accumulated across consecutive workdays, peaked at midweek, and partially dissipated whenever the workweek paused, settling into a rhythm of accumulation and recovery that varied systematically across the workweek itself.

**Day-level conditions that coincide with digital fragmentation**

Having established that digital fragmentation moves with the rhythms of the workweek, we turned to the third question raised in the introduction: which day-level conditions coincide with day-to-day changes in an employee's digital fragmentation?

The two most consequential candidates were the applications that shape how employees coordinate with one another, and the applications that have most recently entered enterprise workflows, namely communication applications[12,13] and generative AI tools[23–25].

We began with communication load, a well-researched topic of study in contemporary knowledge work[13]. Within employees, time spent in communication applications (Teams, Outlook, Slack) showed an inverted-U association with digital fragmentation. Relative to an employee's typical day, fragmentation was higher on days with moderately above-average communication and lower on days when communication occupied most of the workday (within-person linear $\beta = 0.45$, quadratic $\beta = -8.41$, both $P < .001$; Fig. 6A).

To assess whether the lower fragmentation on communication-dominated days reflected only the reduced opportunity to switch across categories; we recomputed the Index after excluding switches within communication applications. The inverted-U finding was robust to this specification (quadratic $\beta = -9.24$, $P < .001$; Supplementary Table S42), indicating that the association was not attributable solely to the reduced opportunity to switch across categories on communication-dominated days. Rather, the data suggest that switching among the remaining non-communication applications was also lower on these days.

Communication load coincided with the interruption of productive work. We measured the productivity interruption rate as the number of transitions from productive to non-productive applications per active hour. We regressed this outcome on the person-mean-centered communication proportion with a control for daily meeting count in a linear mixed-effects model. Work in applications that each company deemed productive was substantially more interrupted on high-communication days (within-person $\beta = 49.5$ interruptions per active hour per unit communication share, $t = 6.34$, $P < .001$; Fig. 6B). The association was robust after adjusting for day of week ($\beta = 49.3$, $P < .001$; Supplementary Table S19).

Results also indicated that communication load predicted fragmentation based on *when* it occurred and *how* much it occurred (Fig. 6C). We measured morning concentration as a Herfindahl-Hirschman Index[32] of category share over the morning window (09:00 to 11:59), with higher values indicating that an employee's pre-noon activity was concentrated in fewer applications. We regressed the afternoon switching rate on morning concentration within person using a linear mixed-effects model with random intercepts for employee and organization (see Methods). On mornings when an employee concentrated their time in a small number of application categories, afternoons carried substantially less switching ($\beta = -54.7$, $t = -17.9$, $P < .001$, 95% CI [-60.7 to -48.7]; Fig. 6C). The association held after controlling for the morning switching rate ($\beta = -33.3$, $P < .001$; Supplementary Table S20), indicating that concentrated mornings preceded consolidated afternoons beyond the persistence of low-switching days, and after adjusting for day of week ($\beta = -53.9$, $P < .001$).

Next, we turned to generative AI tools, the most recent and consequential addition to the digital workday[23,24]. Of the 1,017 employees in the sample, 906 (89.1%) used at least one AI tool during the observation period, with ChatGPT being the most prevalent tool (605 users), followed by Copilot (428), Perplexity (218), Gemini (206), and Claude (50) (Fig. 7A).

AI entered the workday in brief episodes interspersed with other digital applications. Employees turned to AI tools a median of 13 times per day (Fig. 7B), typically only briefly at each visit (Fig. 7C).

Multilevel models that regressed the same-day Fragmentation Index on daily AI-use duration, with day-of-week fixed effects and random intercepts for employee and organization, showed that the Fragmentation Index was higher on days when an employee used AI tools for longer than their own average (within-person $\beta = 0.013$ per minute of AI use, $P < .001$; Fig. 7D).

To characterize the structure of app switching on days when AI was used at-least once, we built a directed weighted application-transition graph at the person-day level[27]. We then computed the modularity (how strongly switching divided into separable groups of related applications) and density (the share of possible application pairs connected by at least one transition) of the subsequent networks. We used these metrics to examine whether switching was organized differently on the days an employee used AI at least once. In separate multilevel models, we regressed network modularity and network density on a person-mean-centered indicator of daily AI use. Person-mean centering compares each employee's AI workdays (when AI was used at least once) with that same employee's non-AI workdays (when AI was not used at all)[33]. Both models included the employee's overall share of AI-use days and day-of-week fixed effects as covariates, with random intercepts for employee and organization (see Methods).

On days when AI was used at least once, an employee's switches packed into more distinct groupings of related applications (modularity, within-person $\beta = 0.028$, $t = 36.6$, $P < .001$) and spanned a smaller share of the possible application pairs (density, $\beta = -0.025$, $t = -41.8$, $P < .001$), even as the overall amount of switching rose ($\beta = 201.3$ more switches per day, $t = 30.5$, $P < .001$). These associations were consistent across all five AI tools (Supplementary Table S43) and disappeared under a within-person placebo relabeling test[34], in which AI-day labels were randomly reassigned within each employee while preserving each employee's overall frequency of AI use, indicating that the associations were tied to the observed timing of AI use rather than to employee-level differences in AI adoption (Supplementary Table S33).

Digital fragmentation rose steadily in the hours leading up to AI use and declined over the hours that followed, tracing a peak at the hour of AI use in the composite Fragmentation Index (Fig. 8A). Each instance of AI use, together with the ten application uses before and after it, formed an episode (n = 880,168). For each episode, we computed five measures over each window: the number of unique applications, the switching rate, the mean duration of a single application use, the share of time in productive applications, and the entropy of application use.

These measures operationalize the breadth, brevity, and unpredictability of switching that the Fragmentation Index captures at the day level into the immediate vicinity of each AI use. Multilevel models estimated the change in each measure from the pre-AI to the post-AI window, with random intercepts for episodes nested within employees (see Methods). In the window following AI use, employees used fewer unique applications ($\beta = -0.060$, $P < .001$; Fig. 8B), switched applications less frequently ($\beta = -0.012$, $P < .001$; Fig. 8C), spent longer in individual applications ($\beta = 531$ ms, $P < .001$; Fig. 8D), devoted a higher share of their time to productive applications ($\beta = 0.0024$, $P < .001$; Fig. 8E), and used applications more predictably ($\beta = -0.018$, $P < .001$; Fig. 8F; Supplementary Table S22). These contrasts hold after adjusting for the day of the week (Supplementary Table S44). The reduction in application breadth and the increase in predictability was consistent across all five AI tools, and the longer dwell times and greater productive share were consistent across four of the five; each tool showed its own characteristic hourly signature in the surrounding hours (Supplementary Fig. S8).

Finally, we examined what employees did after using AI tools (relative to after using other applications) across 56.9 million transitions. To do so, we computed the rate at which each application category followed AI use and compared it with the rate at which that category followed all other applications, expressed as a rate ratio[35] with bootstrap 95% confidence intervals for each category (Supplementary Table S23).

Because every transition lands in exactly one destination category, the categories are interdependent by construction: a higher rate into one category necessarily lowers the rates into others. We therefore treat the rate ratios as a single descriptive profile of post-AI destinations rather than as independent hypothesis tests, and we report the full set of categories together rather than testing each in isolation. Each destination was assigned to one of these categories using the platform's native application labels, which we grouped into a fixed taxonomy of learning platforms, developer applications, browsers, design applications, and Office productivity applications. Employees were disproportionately likely to move from AI tools into learning platforms (RR = 4.04; platform taxonomy) and developer applications (RR = 2.25), and modestly more likely to move into browsers (RR = 1.43). They were less likely to move from AI tools into design applications (RR = 0.55) and, to a smaller degree, into traditional Office productivity applications such as Word and Excel (RR = 0.93). AI use therefore preceded technical and research-oriented work, with administrative work in Office productivity applications somewhat less likely to follow. AI use in this sample marked the points in a workday when employees turned toward investigation, problem-solving, and learning rather than toward Office applications such as Word and Excel.

In a robustness check, we re-derived the destination categories from the platform's native labels in place of our regex-based classifiers. Every shared contrast kept its direction (Supplementary Table S23); the taxonomies differ most for browsers, where the rate attenuates from 1.43 to 1.13 and spans parity under the platform's sparser labels, and for Office productivity, where the reduced rate deepens from 0.93 to 0.32.

**Conclusion**

The modern workday unfolds across digital applications that knowledge workers move through at second-by-second resolution. In prior research, the motion of the contemporary workday has been visible in comparatively short studies of a small number of employees within a small set of organizations. Our analysis of 103 million application events across 1,017 employees and eight organizations, and the development of our Fragmentation Index, brings the picture of digital fragmentation into focus at a more granular time scale, within persons, and across organizations and time.

Researchers have long understood digital fragmentation through one of two perspectives. The first perspective situates digital fragmentation at the level of individual worker, where it has been studied as a function of attention, the tendency to multi-task, and personal habits[4,5,7]. The second perspective situates digital fragmentation at the organizational level, where it has been studied as largely as a result of work design, communication norms, and management practices[36,37].

Our findings reveal that neither of these perspectives in isolation decisively explains where digital fragmentation occurs. The lived experiences of employees, including how they structure their mornings and workweeks, explains a sizeable component of the changes in digital fragmentation that we observe across workdays. However, the largest share of variation in digital fragmentation is explained by variation in digital fragmentation *within* the same employee over time. The same individual is more heavily fragmented on some days, and not others, and this variability is partially attributable to employees' work activities and rest periods. Specifically, digital fragmentation increases across the workweek, peaks midweek, and resets after breaks, and varies with day-to-day conditions such as communication demands and AI use. This interpretation is consistent with research showing substantial within-person behavioral variability across days, echoing the distinction between relatively stable traits and more variable psychological states[38], and with evidence that day-level conditions shape the temporal organization of work[9,39]. The workday therefore constitutes a substantive level of analysis for understanding digital fragmentation: when work occurs, and under what daily conditions, helps determine how fragmented digital work becomes.

Building upon decades of research on attentional residue, our findings indicate that the consequences of one work period can persist into the next[8,9]. Indeed, our results suggest that protecting the first hours of work may have disproportionate value because such protection occurs before fragmentation has accumulated, and interruption costs have propagated across subsequent tasks.

Another question our findings raise concerns the pacing of demanding work across the workweek. Fragmentation accumulates as the week goes on, peaks at midweek, and eases at the start and the end of the week. If demanding work is concentrated at the midweek peak, it may coincide with fragmentation that is already at its highest. Thus, midweek communication demands are a natural focal point for future intervention research on digital fragmentation.

Finally, the weekend and holiday patterns connect digital fragmentation to the recovery literature, which has largely measured recovery through self-reports of psychological detachment, relaxation and resource restoration[40,41]. Our results reveal a behavioral signature of this process: fragmentation accumulated across consecutive workdays and partially reset after weekends and national holidays. After-hours communication and weekend work may therefore erode the recovery on which the following workweek depends, not only by reducing subjective detachment (which previous research has shown), but by carrying fragmentation forward into the behavioral structure of subsequent workdays.

Commentators have warned that AI tools could add yet another distraction, intensify cognitive load, and fragment the workday even further[42]. Our findings challenge this view. On days when employees use AI, total switching rises but becomes more structured; in the moments immediately after each AI use, application use consolidates. Employees move through fewer applications, dwell longer in each application, and follow AI use with greater time spent in research and technical work rather than a return to administrative tasks. This finding connects to a growing body of experimental research showing that generative AI improves the quality and speed of work on cognitively demanding tasks[22,25].

Three limitations bound these findings. First, our data capture behavior at scale, not the attention, intent, or output behind it[42]. Therefore, the association between AI use and a reorganized workday enables more than one interpretation: AI use may reorganize the day, employees may turn to AI on days that are already consolidated, or both. Distinguishing between these accounts will require pairing this kind of data with experience sampling and experiments that manipulate AI use directly. Second, whether an application counts as productive follows each organization's own classification, the standard operationalization in workforce analytics[12]; however, because productivity also encompasses task quality, cognitive effort, and subjective engagement[20,43], pairing these labels with task-level outcomes, self-reports, and direct measures of effort would meaningfully expand the measure. Third, our eight organizations are based in India, where local norms can shape workday boundaries and after-hours availability[44,45]. India is the world's leading exporter of digital knowledge work[46,47]. Knowledge workers in India operate within global corporate networks on the same software ecosystems used across the world. This leads us to attribute what we observe to these common tools rather than to local conditions. Cross-regional research will establish which of these regularities are universal and which reflect cultural differences.

Together, these results suggest that digital fragmentation varies more with the day and the person than with the organization, and that application use becomes more consolidated in the moments after AI use. We find that digital fragmentation varies more by the day and the employee than by the organization, so individual work habits and daily routines [8,9,39] are more promising targets of intervention than firm-wide policies alone[36,37]. We also show that digital fragmentation appears to accumulate across workdays and reset over weekends and holidays. These findings suggest that *when* an intervention occurs may matter as much as intervening in the first place. And finally, because generative AI is rapidly entering the workplace and coincides with more consolidated rather than more fragmented application use, AI may be part of the solution as opposed to a contributor to the problem of digital fragmentation.

Using, to our knowledge, the largest and most granular record of digital work yet assembled, comprising 103 million second-by-second application events from 1,017 knowledge workers across eight organizations, this study identifies the workday as the central unit of digital fragmentation. Across 62,693 person-days and 56.9 million application switches, fragmentation rose with communication demands and across the work week, then reset after weekends and holidays. Generative AI use also occurred on more fragmented days, but the period following AI use was marked by narrower, longer, and more predictable application use. These findings identify the workday as a key level for understanding and intervening on digital fragmentation and suggest that AI may help structure fragmented work rather than merely intensify it.

**References**

1. Drucker, P. F. Knowledge-Worker Productivity: The Biggest Challenge. *California Management Review* **41**, 79–94 (1999).
2. González, V. M. & Mark, G. 'Constant, constant, multi-tasking craziness': managing multiple working spheres. in *Proceedings of the SIGCHI conference on human factors in computing systems* 113–120 (Association for Computing Machinery, New York, NY, USA, 2004).
3. Murty, R. N., Dadlani, S. & Das, R. B. How much time and energy do we waste toggling between applications? *Harvard Business Review* (2022).
4. Rubinstein, J. S., Meyer, D. E. & Evans, J. E. Executive control of cognitive processes in task switching. *Journal of Experimental Psychology: Human Perception and Performance* **27**, 763–797 (2001).
5. Monsell, S. Task switching. *Trends in Cognitive Sciences* **7**, 134–140 (2003).
6. Mark, G., Gudith, D. & Klocke, U. The cost of interrupted work: more speed and stress. in *Proceedings of the SIGCHI conference on human factors in computing systems* 107–110 (Association for Computing Machinery, New York, NY, USA, 2008).
7. Ophir, E., Nass, C. & Wagner, A. D. Cognitive control in media multitaskers. *Proceedings of the National Academy of Sciences* **106**, 15583–15587 (2009).
8. Leroy, S. & Schmidt, A. M. The effect of regulatory focus on attention residue and performance during interruptions. *Organizational Behavior and Human Decision Processes* **137**, 218–235 (2016).
9. Leroy, S. Why is it so hard to do my work? The challenge of attention residue when switching between work tasks. *Organizational Behavior and Human Decision Processes* **109**, 168–181 (2009).
10. Mark, G., Iqbal, S. T., Czerwinski, M., Johns, P. & Sano, A. Neurotics can't focus: an in situ study of online multitasking in the workplace. in *Proceedings of the 2016 CHI conference on human factors in computing systems* 1739–1744 (Association for Computing Machinery, New York, NY, USA, 2016).
11. Mark, G. *Attention Span: A Groundbreaking Way to Restore Balance, Happiness and Productivity*. (Harlequin, 2023).
12. Yang, L. *et al.* The effects of remote work on collaboration among information workers. *Nature Human Behaviour* **6**, 43–54 (2022).
13. DeFilippis, E., Impink, S. M., Singell, M., Polzer, J. T. & Sadun, R. The impact of COVID-19 on digital communication patterns. *Humanities and Social Sciences Communications* **9**, 1–11 (2022).
14. Polzer, J. T. The rise of people analytics and the future of organizational research. *Research in Organizational Behavior* **42**, 100181 (2022).
15. Czerwinski, M., Horvitz, E. & Wilhite, S. A diary study of task switching and interruptions. in *Proceedings of the SIGCHI conference on human factors in computing systems* 175–182 (Association for Computing Machinery, New York, NY, USA, 2004).
16. Mark, G. *et al.* Email duration, batching and self-interruption: patterns of email use on productivity and stress. in *Proceedings of the 2016 CHI conference on human factors in computing systems* 1717–1728 (Association for Computing Machinery, New York, NY, USA, 2016).

17. Jin, J. & Dabbish, L. A. Self-interruption on the computer: a typology of discretionary task interleaving. in *Proceedings of the SIGCHI conference on human factors in computing systems* 1799–1808 (Association for Computing Machinery, New York, NY, USA, 2009).
18. Yeykelis, L., Cummings, J. J. & Reeves, B. Multitasking on a single device: arousal and the frequency, anticipation, and prediction of switching between media content on a computer. *Journal of Communication* **64**, 167–192 (2014).
19. Meyer, A. N., Barton, L. E., Murphy, G. C., Zimmermann, T. & Fritz, T. The work life of developers: activities, switches and perceived productivity. *IEEE Transactions on Software Engineering* **43**, 1178–1193 (2017).
20. Mark, G., Iqbal, S., Czerwinski, M. & Johns, P. Focused, aroused, but so distractible: temporal perspectives on multitasking and communications. in *Proceedings of the 18th ACM conference on computer supported cooperative work & social computing* 903–916 (Association for Computing Machinery, Vancouver, BC, Canada, 2015).
21. Coviello, D., Ichino, A. & Persico, N. Time allocation and task juggling. *American Economic Review* **104**, 609–623 (2014).
22. Bick, A., Blandin, A. & Deming, D. J. The rapid adoption of generative AI. *Management Science* https://doi.org/10.1287/mnsc.2025.02523 (2026) doi:10.1287/mnsc.2025.02523.
23. Brynjolfsson, E., Li, D. & Raymond, L. R. Generative AI at work. *Quarterly Journal of Economics* **140**, 889–942 (2025).
24. Dell'Acqua, F. *et al.* Navigating the jagged technological frontier: field experimental evidence of the effects of artificial intelligence on knowledge worker productivity and quality. *Organization Science* **37**, 403–423 (2026).
25. Noy, S. & Zhang, W. Experimental evidence on the productivity effects of generative artificial intelligence. *Science* **381**, 187–192 (2023).
26. Roehrick, K., Vaid, S. S. & Harari, G. M. Situating Smartphones in Everyday Life: Personality Traits and Situational Context Shape Momentary Smartphone Use. *Journal of Personality and Social Psychology* (2023).
27. Turner, G. *et al.* Relating smartphone 'App-journeys' to mental health: A novel network analysis approach. Preprint at https://osf.io/preprints/psyarxiv/uytfr_v1/ (2026).
28. Shannon, C. E. A mathematical theory of communication. *Bell System Technical Journal* **27**, 379–423 (1948).
29. Newman, M. E. J. *Networks: An Introduction*. (Oxford University Press, Oxford, 2010).
30. Bryk, A. S. & Raudenbush, S. W. Application of hierarchical linear models to assessing change. *Psychological bulletin* **101**, 147 (1987).
31. Bolger, N., Davis, A. & Rafaeli, E. Diary Methods: Capturing Life as it is Lived. *Annual review of psychology* **54**, 579–616 (2003).
32. Rhoades, S. A. The Herfindahl-Hirschman index. *Federal Reserve Bulletin* **79**, 188–189 (1993).
33. Yaremych, H. E., Preacher, K. J. & Hedeker, D. Centering categorical predictors in multilevel models: Best practices and interpretation. *Psychological Methods* **28**, 613–630 (2023).
34. Winkler, A. M., Ridgway, G. R., Webster, M. A., Smith, S. M. & Nichols, T. E. Permutation inference for the general linear model. *Neuroimage* **92**, 381–397 (2014).
35. Rothman, K. J., Greenland, S. & Lash, T. L. *Modern Epidemiology*. (Lippincott Williams & Wilkins, 2008).

36. Parker, S. K., Morgeson, F. P. & Johns, G. One hundred years of work design research: Looking back and looking forward. *Journal of Applied Psychology* **102**, 403–420 (2017).
37. Humphrey, S. E., Nahrgang, J. D. & Morgeson, F. P. Integrating motivational, social, and contextual work design features: A meta-analytic summary and theoretical extension of the work design literature. *Journal of Applied Psychology* **92**, 1332–1356 (2007).
38. Fleeson, W. Moving personality beyond the person-situation debate: The challenge and the opportunity of within-person variability. *Current Directions in Psychological Science* **13**, 83–87 (2004).
39. Whillans, A., Perlow, L. & Turek, A. Experimenting during the shift to virtual team work: Learnings from how teams adapted their activities during the COVID-19 pandemic. *Information and Organization* **31**, 100343 (2021).
40. Sonnentag, S., Völker, J. & Wehrt, W. Good and bad days at work: A descriptive review of day-level and experience-sampling studies. *Journal of Organizational Behavior* **n/a**, (2024).
41. Fritz, C., Sonnentag, S., Spector, P. E. & McInroe, J. A. The weekend matters: Relationships between stress recovery and affective experiences. *J Organ Behavior* **31**, 1137–1162 (2010).
42. Salganik, M. J. *Bit by Bit: Social Research in the Digital Age*. (Princeton University Press, Princeton, 2018).
43. Deng, T. *et al.* Measuring smartphone usage and task switching with log tracking and self-reports. *Mobile Media & Communication* **7**, 3–23 (2019).
44. Gelfand, M. J. & others. Differences between tight and loose cultures: a 33-nation study. *Science* **332**, 1100–1104 (2011).
45. Belkin, L. Y., Becker, W. J. & Conroy, S. A. The invisible leash: the impact of organizational expectations for email monitoring after-hours on employee resources, well-being, and turnover intentions. *Group & Organization Management* **45**, 709–740 (2020).
46. Dossani, R. & Kenney, M. The next wave of globalization: relocating service provision to India. *World Development* **35**, 772–791 (2007).
47. Bloom, N., Genakos, C., Sadun, R. & Van Reenen, J. Management practices across firms and countries. *The Academy of Management Perspectives* **26**, 12–33 (2012).

## Methods

### Data source

We analyzed application-use records from 1,017 employees in eight Indian organizations, collected between November 2025 and January 2026. The records were collected by Flowace, a commercial workforce-analytics service used by all eight participating organizations.

Flowace's software runs on each employee's work on their computer and writes a record every time the employee changes the application that is in active use. For each record, Flowace stores the application name, the time the application becomes active, the time it stops being active, the window title, and the URL, where applicable. Each record is time-stamped within a fraction of a second. Flowace also stores a classification of every application as "productive" or "non-productive" which is a classification that is supplied by each employer for use within Flowace's own reporting interface, and we received this classification with the records we analyzed.

The eight participating organizations span consulting, technology, financial services, and professional services. The complete set of records contained 102,798,453 application uses, observed over a median of 58 workdays per employee. The Harvard University Committee on the Use of Human Subjects approved the protocol (IRB 25-0163), and Flowace and the participating organizations transferred the records to the research team under a data-use agreement. Per-organization composition and observation windows appear in Supplementary Table S3. On average, each organization contributed 12.8 million records from 127 employees (range 24 to 239 employees per organization), and each employee contributed roughly 101,100 records.

### Sample

The analytic unit is the person-day, defined as one employee's application use on a single calendar day ranging from 12:01 AM to 11:59 PM local time (IST). Each employee is identified by a key that combines the organization and the employee identifier within the organization, and every per-employee statistic in this paper counts unique employees by using that key. Descriptive statistics for both analytic samples appear in Supplementary Table S2.

The records originally arrived from nine organizations. We dropped one organization from the analytic sample because the records from that organization were entirely self-reported timesheet entries  with no automatically recorded application use, making the data incomparable to the other data that we received. The eight remaining organizations supplied the 1,017 employees described above. For the fragmentation analyses, we further restricted the sample to employees observed for at least ten active workdays, the minimum we considered adequate for the within-person centering, lag-1 carryover, and day-of-week analyses reported in Section 2 of the main text. This restriction excluded 39 employees and left an analytic sample for our fragmentation analyses of 978 employees and 52,882 person-days. The analyses of generative AI use reported in Section 3 retained all 1,017 employees so as not to drop sporadic AI users, and the 39-employee discrepancy is noted wherever sample sizes are reported in text.

### Data Cleaning

We removed application logs that lasted zero seconds, and those that lasted longer than four hours. The four-hour cap is well above the 99th percentile of duration in our records (259 seconds) and removes a small residual of implausibly long single-application records that most likely originate from workstations left unattended with one application left open in the foreground. Next, we removed four kinds of pseudo-records that Flowace writes into the raw records, but that do not correspond to actual application use. These were Flowace's own internal status records, idle-time placeholders that Flowace inserts when no user input is detected, self-reported timesheet entries that the employee writes into the platform after the fact and that carry hour-scale durations, and meeting placeholders derived from each employee's connected work calendar. The meeting placeholders were retained separately as a daily count of scheduled meetings for use in the communication-demand regression.

When two consecutive records carry the same application name (for example, when an employee moved between two Word documents and produced two records both labeled Microsoft Word), we treated the two records as one continuous use of that application rather than as a switch from one application to another. The 56,943,860 application switches analyzed and reported in the main manuscript are switches from one application to a different application.

Data cleaning decisions and how many employees and corresponding observations were excluded are cataloged in Supplementary Table S5. Sensitivity of the main findings to each of these cleaning decisions is reported in Supplementary Tables S25 to S28.

**Session definition and preprocessing**

A session is a continuous bout of application use. We marked the start of a new session whenever the next application use began more than five minutes after the end of the previous one, and we treated a session as analytically valid if it included at least three distinct applications. The five-minute threshold separates within-task transitions from between-task transitions in line with the framework introduced by Gonzalez and Mark and matches the convention used by other commercial workforce-analytics services. Because the threshold is conventional rather than empirically derived, we report sensitivity at 1, 10, 15, and 30 minutes in Supplementary Table S28. Across the records, the median application event lasted 2.15 seconds, and the median bout within productivity applications lasted 5.74 seconds (Supplementary Table S4).

We segmented each person-day's events into sessions using a five-minute inter-event gap rule. A new session began whenever the time between consecutive events exceeded five minutes or could not be computed (the first event of each person-day, by construction).

We retained sessions for analysis only if they contained at least three distinct applications, the minimum required for the within-session graph and entropy measures (transition entropy, edge count, density, modularity) to be well-defined on every session contributing to the Fragmentation Index. 34,573 of the 358,727 sessions detected across the eight participating organizations (9.64%) failed this criterion and were excluded, with a median duration of 91.7 seconds, compared with a median duration of 50.6 minutes for retained sessions.

Because the filter is applied at the session level, it affects all analyses computed from within-session structure, including the Fragmentation Index and its three-level variance decomposition, the network-graph rows of Table 2 (transition entropy, modularity, density), the application destination rate ratios, the descriptive session statistics in Supplementary Table S4, and the communication-fragmentation interruption analyses in Section 2.

Analyses computed at the day level from individual application uses are not affected by this filter, including the productive-time proportion, the distraction rate, the recovery rate, the peri-event panels around generative AI use (Figure 5A-E across 880,168 episodes), the AI daily descriptive statistics (Figure 4A-C), and the AI tool heterogeneity comparisons. Supplementary Tables S29 to S31 report a sensitivity analysis of the affected findings under minimum-session thresholds of 1, 2, 3 (the value used in the main text), 5, and 10 distinct applications, with the tolerance rules and decision criteria documented in the supplementary robustness log (Supplementary Tables S25 to S32). Across the corpus, the median application event lasted 2.15 seconds, and the median productivity-application bout was 5.74 seconds.

**Fragmentation Index**

We constructed a composite Fragmentation Index for each person-day from three indicators that capture the volume and the structure of application switching. Transition entropy is a Shannon entropy measure[44] of the unpredictability of an employee's application switches, computed over the edge-weight distribution of the within-day application transition graph. Network edge count is the number of distinct ordered application pairs used during the day, indexing the breadth of switching pathways. Mean session duration is the average length of time an employee spent in an application before switching, reverse-scored so that higher values reflect greater fragmentation.

Each indicator was z-scored across person-days, and a principal-components analysis was fit on the three z-scored indicators (centered, unscaled). The first principal component (PC1) explained 66.8% of the joint variance, and component loadings were dominated by transition entropy and network edge count (0.70 each), with a smaller contribution from reverse-scored mean session duration (−0.15), after a sign flip ensuring that higher PC1 values corresponded to greater fragmentation. We retained PC1 as the Fragmentation Index used throughout the manuscript. The component loadings appear in Supplementary Table S10, and each measure is defined in Supplementary Table S1.

**Variance decomposition**

We partitioned variance in the Fragmentation Index and in each component metric into three nested levels — organizations, persons within organizations, and days within persons — using random-intercept null models fit by restricted maximum likelihood[49]. Variance components were converted to intraclass correlation coefficients by dividing each variance term by the total variance. Component-level partitions are reported in Table 1; for productive-time proportion, 85.6% of variance was between employees and organizations combined. Week-over-week stability of the Fragmentation Index within employees was ICC = 0.71. The full decomposition for the Index and each component appears in Supplementary Table S11.

**Auxiliary day-level measures**

For each person-day, we computed network and productivity metrics directly from the application-transition graph and the event stream.

**Network metrics.** We constructed a directed weighted graph in which nodes were applications and edges were ordered transitions weighted by frequency, using the igraph package[45]. We extracted the number of nodes and edges, edge density, reciprocity, global transitivity, modularity from a Louvain community-detection algorithm[46] applied to the undirected collapse of the graph, and the number of detected communities.

**Productivity metrics.** From each person-day we computed the proportion of time spent in productivity applications, the rate of productivity interruptions (productivity-to-non-productivity transitions per active hour), the recovery rate (the inverse rate at which interrupted productivity bouts resumed), and the total number of application switches (mean = 968 switches per person-day on workdays). The productivity classification is documented in Supplementary Table S7.

**Communication load.** Communication intensity was operationalized at the person-day level as the proportion of application events classified as communication tools, using a project-specific regular-expression taxonomy applied to application names and URLs (Supplementary Table S6). For models requiring a within-person decomposition, we additionally constructed person-mean-centered versions of all communication and productivity predictors[47].

**Morning concentration.** For analyses of morning-to-afternoon dynamics, we computed a Herfindahl-Hirschman Index[48] of category share over the morning window of each workday, indexing how concentrated an employee's pre-noon activity was on a single category of work.

### Generative AI tool detection and episode windowing

We identified generative AI use through a regular-expression rule set that matched application names, executable names, and URLs to five widely used generative AI tools (ChatGPT, Microsoft Copilot, Perplexity, Google Gemini, and Anthropic Claude). The full rule set appears in Supplementary Table S8. We report counts of AI users in two ways. The first counts an employee as a user of a given AI tool only if the employee accessed that tool through its primary path, for example chatgpt.com for ChatGPT. The second additionally counts employees who accessed the tool through embedded surfaces inside other applications, command-line interfaces, and other secondary paths. Both sets of counts are provided in Supplementary Table S9. Of the 1,017 employees in the AI analytic sample, 906 (89.1%) used at least one of the five tools during the observation period.

For the peri-event analyses, we defined an AI episode as the ten application uses immediately preceding a single use of an AI tool together with the ten immediately following it, restricted to the same employee-day. The records contained 880,168 such episodes. For each episode, we computed five quantities separately for the preceding and following ten uses, namely the number of distinct applications, the rate of application switching, the mean duration of a single application use, the duration-weighted share of time in productive applications, and the Shannon entropy of the application distribution. The resulting window contrasts are reported in Supplementary Tables S22 and S44.

### Multilevel models

All multilevel models were fit in R using lme4 with Satterthwaite degrees of freedom from lmerTest[50], the BOBYQA optimizer, and a maximum of 50,000 (or 100,000, where convergence required) function evaluations. The estimator and random-effects structure for every model family are cataloged in Supplementary Table S36.

**Communication-load models.** We estimated the within-person association between communication intensity and fragmentation by regressing the productivity interruption rate on person-mean-centered communication proportion and meeting count, with random slopes for communication proportion and random intercepts for tenant. Where the random slope produced a singular fit, models were re-estimated with random intercepts only. Between-person counterparts added person-mean terms and a fully nested random structure. Morning-to-afternoon carryforward was estimated by regressing afternoon fragmentation on morning communication share and on the morning concentration index. Full estimates appear in Supplementary Tables S19 and S42.

**AI within-person models.** We compared AI-use and non-AI days within the same employee, regressing each outcome on person-mean-centered AI-use, the employee's overall share of AI-use days, and a fixed-effects factor for day of week, with random intercepts for tenant and employee. Models were fit separately for each of seven outcomes (productive-time proportion, transition entropy, modularity, density, distraction rate, recovery rate, and the Fragmentation Index). Full estimates appear in Supplementary Table S21.

**Day-of-week and weekend models.** These models estimated the diurnal U-shape in fragmentation and a weekend-by-hour interaction, with a comparable random-effect structure. The corresponding estimates appear in Supplementary Tables S13 to S15.

**Holiday event study models.** Using a reference calendar of 37 Indian national holidays spanning 2025-2026, five of which fell within the observation window, we estimated fragmentation over the five workdays before and after each holiday, treating the holiday calendar as external to day-level digital behavior because holidays are fixed by the national calendar in advance and cannot respond to how employees use their applications. We aligned each employee's within-person (z-scored) Fragmentation Index to the workdays surrounding each holiday; day 0 is the holiday itself and is not a workday. Each of the ten surrounding workdays entered a linear mixed model as an indicator term, with all workdays outside any holiday window serving as the reference category, and with random intercepts for employees nested within organizations. Because the November-to-January calendar is dense with holidays, a single date can fall near two holidays at once; each date was therefore assigned to its nearest holiday only, and any date falling within five workdays of two or more holidays was excluded, so that the days at the edges of the window are estimated only from isolated holidays (Supplementary Table S16).

We checked that these estimates reflect the holidays themselves rather than calendar position in two ways. First, we re-estimated the model with day-of-week fixed effects; the eve-of-holiday and first-day-back estimates persisted (Supplementary Table S16). Second, we constructed placebo holidays: 1,000 sets of 37 pseudo-holiday dates, each set matched to the real holidays' day-of-week distribution and restricted to dates at least five workdays from any real holiday. For each set we computed the same event-day profile of the within-person Index, and took the 2.5th to 97.5th percentiles of the 1,000 placebo profiles as a null band for each event day. The placebo profiles show no dip around pseudo-holidays, and the observed holiday estimates fall outside the null band (Supplementary Table S17).

**AI-destination rate ratios.** Each of the 56,943,860 application transitions was classified once by the category of its destination application, so that the destination categories are mutually exclusive and exhaustive and form a single multinomial distribution per source type. We compared the distribution of destinations conditional on an AI source against the distribution conditional on all non-AI sources, summarizing each category with a rate ratio, RR = P(destination | source = AI) / P(destination | source ≠ AI). Because the category shares sum to one, the rate ratios are interpreted jointly as a single redistribution profile rather than as independent tests of separate outcomes; an omnibus chi-square test on the source-by-destination contingency table evaluates whether the two distributions differ overall (Supplementary Table S24). Ninety-five percent confidence intervals were obtained from a person-day block bootstrap (1,000 resamples; the resampling unit is the employee-day, so all transitions contributed by an employee on a single day enter or leave a resample together), which accounts for the dependence among transitions from the same employee-day without modeling assumptions. The rate ratios with bootstrap confidence intervals appear in Supplementary Table S23.

**Robustness analyses**

We report seven robustness analyses, with tolerance rules and decision criteria documented in the supplementary robustness log (Supplementary Tables S25 to S32). We re-estimated the main-text models after each of the following changes to the data construction. We relaxed the four-hour duration cap. We relaxed the ten-day employee inclusion filter. We retained the calendar-derived meeting placeholders in the cleaned records. We retained the self-reported timesheet entries in the cleaned records. We varied the five-minute session-gap threshold across 1, 10, 15, and 30 minutes. We re-estimated Table 2 using an analyst-defined regular-expression list of productive applications in place of the employer-supplied classification. We re-computed destination rate ratios over narrower application sets, namely Word and Excel alone for Office productivity, classic integrated development environments alone for developer work, and Photoshop, Illustrator, and Figma alone for design. Results from each analysis appear in Supplementary Tables S25 to S32. We additionally recomputed all destination rate ratios comparing our analyst-defined regular-expression taxonomy against the platform's native categories; the direction of every shared contrast held, with the browser ratio attenuating to non-significance under the platform's sparser labels (1.43 to 1.13) and the Office-productivity ratio moving further below parity (0.93 to 0.32) (Supplementary Table S23).

We additionally estimated the AI-day effects separately for each of the five AI tools to assess per-tool consistency (Supplementary Table S43). We conducted a placebo analysis in which AI-day labels were randomly reassigned within each employee while preserving the employee's marginal frequency of AI use (Supplementary Table S33). For the within-person AI models, we conducted selection-on-observables sensitivity tests, adjusting for the available day-level covariates and assessing coefficient stability (Supplementary Table S35).

**Software and data availability**

The behavioral trace data analyzed in this study cannot be made publicly available because they contain proprietary workforce-analytics records provided under data-use agreements with participating organizations that prohibit redistribution, and because the granularity of second-by-second application-use sequences creates a reidentification risk that cannot be fully mitigated through anonymization or aggregation. Summary statistics, variance decomposition tables, and all model outputs necessary to evaluate the reported findings are provided in the Supplementary Materials. Analysis code that documents the procedures used to generate these outputs was provided with the manuscript submission.

**Methods references**

48. Shannon, C. E. A mathematical theory of communication. *Bell Syst. Tech. J.* **27**, 379–423 (1948).

49. Csardi, G. & Nepusz, T. The igraph software package for complex network research. *InterJournal Complex Systems*, 1695 (2006).

50. Blondel, V. D., Guillaume, J.-L., Lambiotte, R. & Lefebvre, E. Fast unfolding of communities in large networks. *J. Stat. Mech.* **2008**, P10008 (2008).

51. Enders, C. K. & Tofighi, D. Centering predictor variables in cross-sectional multilevel models: a new look at an old issue. *Psychol. Methods* **12**, 121–138 (2007).

52. Hirschman, A. O. The paternity of an index. *Am. Econ. Rev.* **54**, 761–762 (1964).

53. Bates, D., Mächler, M., Bolker, B. & Walker, S. Fitting linear mixed-effects models using lme4. *J. Stat. Softw.* **67**, 1–48 (2015).

54. Kuznetsova, A., Brockhoff, P. B. & Christensen, R. H. B. lmerTest package: tests in linear mixed effects models. *J. Stat. Softw.* **82**, 1–26 (2017).

55. R Core Team. *R: A Language and Environment for Statistical Computing* (R Foundation for Statistical Computing, Vienna, 2023).

56. Barrett, T., Dowle, M., Srinivasan, A., Gorecki, J., Chirico, M. & Hocking, T. *data.table: Extension of `data.frame`*. R package version 1.17.6 (2025).

57. Richardson, N. et al. *arrow: Integration to 'Apache' 'Arrow'*. R package version 20.0.0.2 (2025).

58. Brooks, M. E. et al. glmmTMB balances speed and flexibility among packages for zero-inflated generalized linear mixed models. *R J.* **9**, 378–400 (2017).

59. Viechtbauer, W. Conducting meta-analyses in R with the metafor package. *J. Stat. Softw.* **36**, 1–48 (2010).

60. Wickham, H. *ggplot2: Elegant Graphics for Data Analysis* (Springer-Verlag, New York, 2016).

**Acknowledgments:** We thank Chanelle Scheffer, Jenny Lin and other members of the Harvard community for providing valuable feedback on early versions of the manuscript.

**Funding:** This research was not funded by any external sources.

**Author contributions:**

Conceptualization: SSV, AVW

Methodology: SSV

Investigation: SSV

Visualization: SSV, AVW

Funding acquisition: AVW

Project administration: SSV, AVW

Supervision: AVW

Writing – original draft: SSV, AVW

Writing – review & editing: SSV, AVW

**Diversity, equity, ethics, and inclusion [optional]:** This study analyzes passively recorded digital trace data collected by Flowace, a workforce analytics platform deployed by participating organizations as part of their standard operations. All data were fully anonymized prior to transfer to the research team; no individually identifying information was available at any stage of the analysis. The research protocol was reviewed and approved by IRB 25-0163. While the sample spans multiple industries and organizational structures, it does not represent all segments of the Indian workforce, nor workers in the informal economy, manufacturing, or service sectors where digital trace methods are inapplicable. The generalizability of these findings to workers outside the knowledge economy should not be assumed.

**Competing interests:** Flowace provided de-identified platform records under a data-use agreement but provided no funding or other commercial support for this research, had no role in the study design, analysis, or interpretation, and had no discretion over what could be published. SSV holds equity in a private company (not related to Flowace) that uses AI to create enterprise digital twins. AVW holds equity and advises the company.

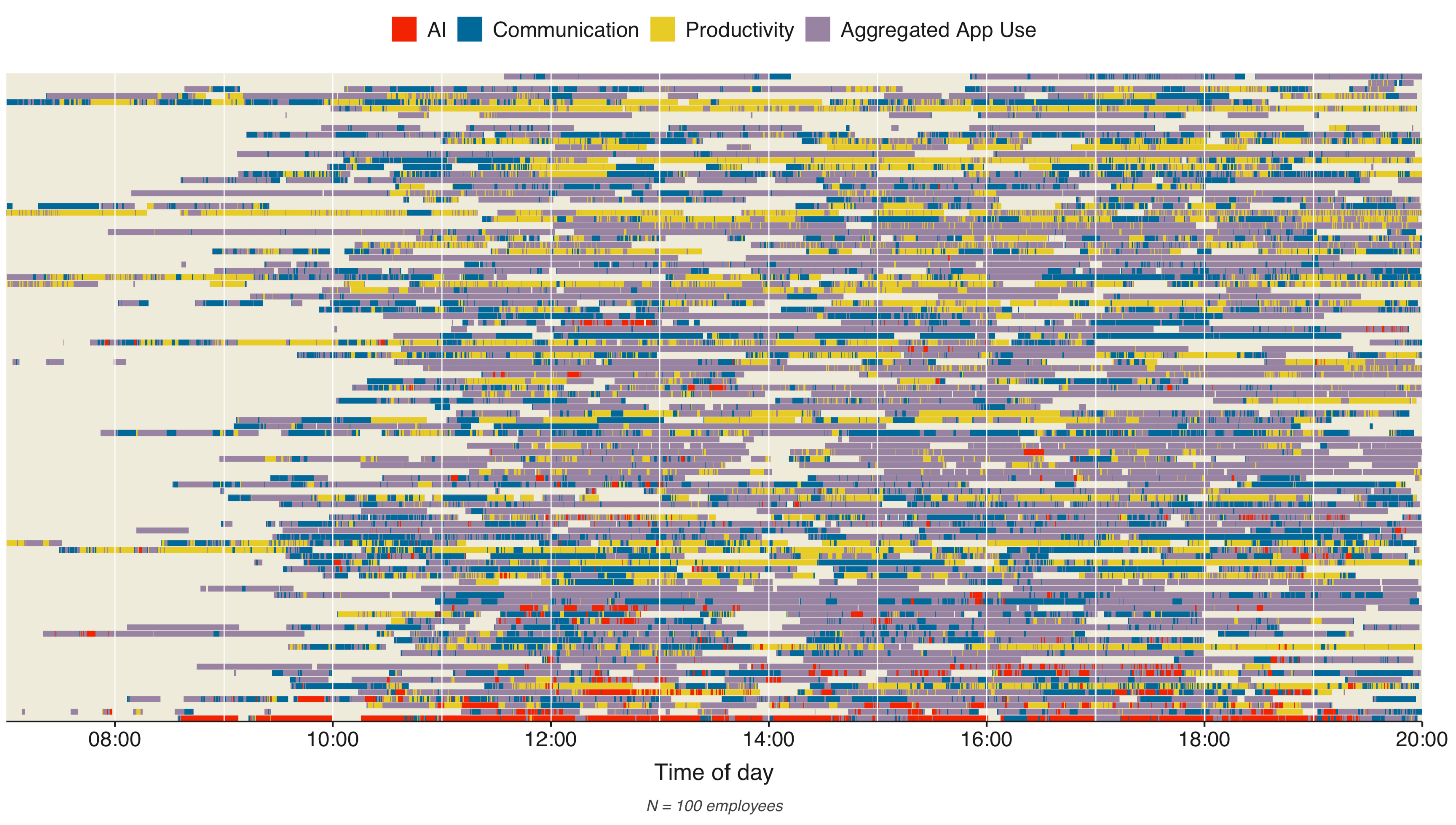

**Fig. 1 | The digital workday is a fast and varied succession of brief application uses, with employees interleaving productivity, communication, and generative AI tools across the hours of the day.** Each horizontal row depicts the time-ordered sequence of application categories used by a single employee across one workday ($N$ = 100 employees). The four most prevalent application categories are colored: generative AI (red), communication (blue), productivity (yellow), and aggregated app use (lavender). Periods without digital activity appear as the cream background. Rows are sorted from least AI use (top) to greatest AI use (bottom).

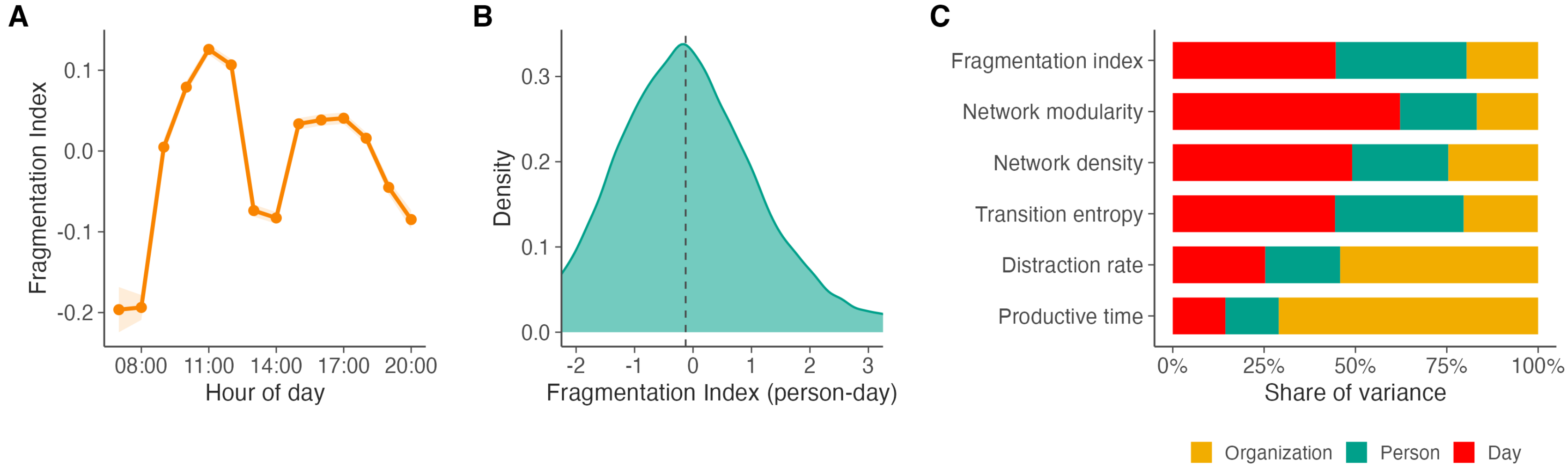

**Fig. 4 | The Fragmentation Index rises and falls within the workday, distributes unimodally across person-days, and splits its variance roughly evenly between the person and the day. a.** Mean Fragmentation Index across the hours of the workday. The index rises from a morning trough to a mid-morning peak near 11:00, dips at midday, climbs again to a smaller mid-afternoon peak near 17:00, and declines toward the close of the day. **b.** Distribution of the Fragmentation Index at the person-day level, with the dashed line marking the sample mean. The distribution is unimodal with a longer right tail capturing the more heavily fragmented end. **c.** Three-level variance decomposition for six digital work metrics. Each bar partitions total variance into day-level (red), person-level (teal), and organization-level (orange) components. The Fragmentation Index splits nearly evenly between the day (44.6%) and the person (35.8%), distinguishing it from distraction rate and productive time, which are overwhelmingly person-level.

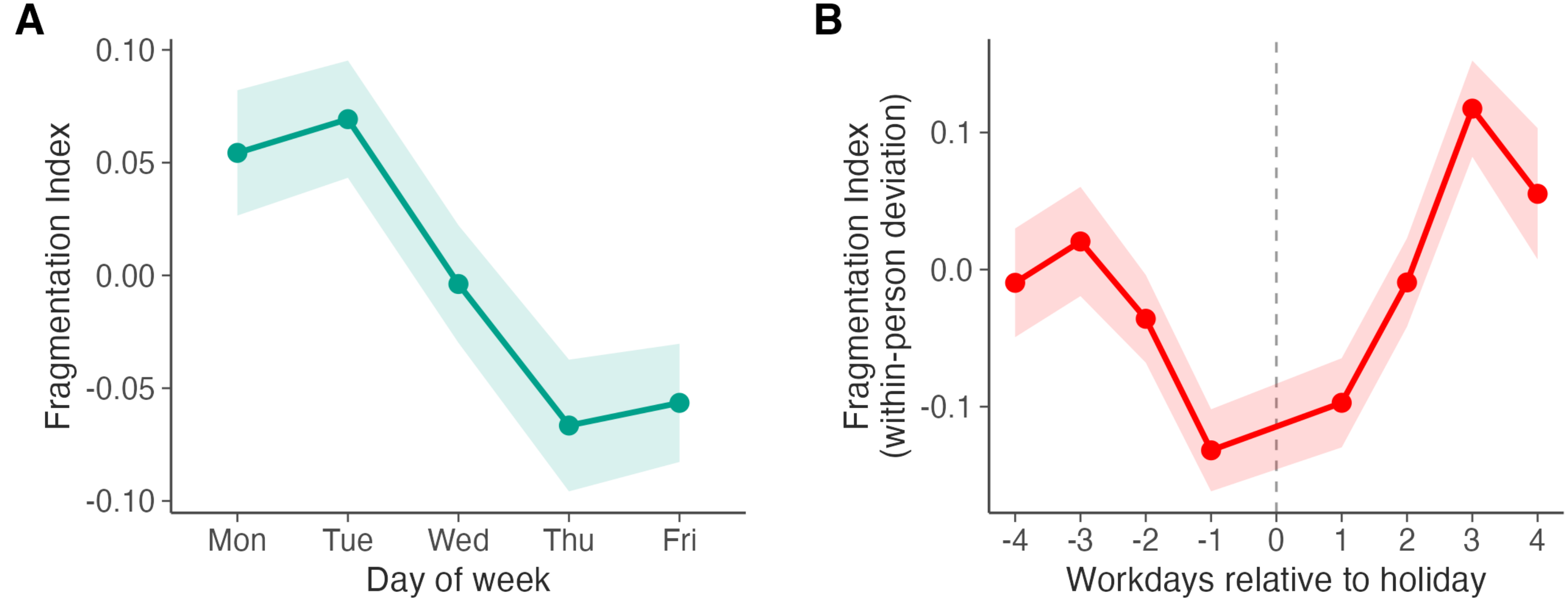

**Fig. 5 | Digital fragmentation follows the rhythm of the workweek and partially resets after holidays. a.** Mean Fragmentation Index by day of the workweek. Fragmentation peaks on Tuesday and declines toward Friday, with the shaded ribbon indicating the 95% confidence interval. **b.** Within-person Fragmentation Index across the five workdays before and after each in-window Indian national holiday (five of the 37-holiday reference calendar). Day 0 marks the holiday itself. Fragmentation declines steadily across the workdays preceding a holiday, drops sharply on the holiday, and partially rebounds in the workdays that follow.

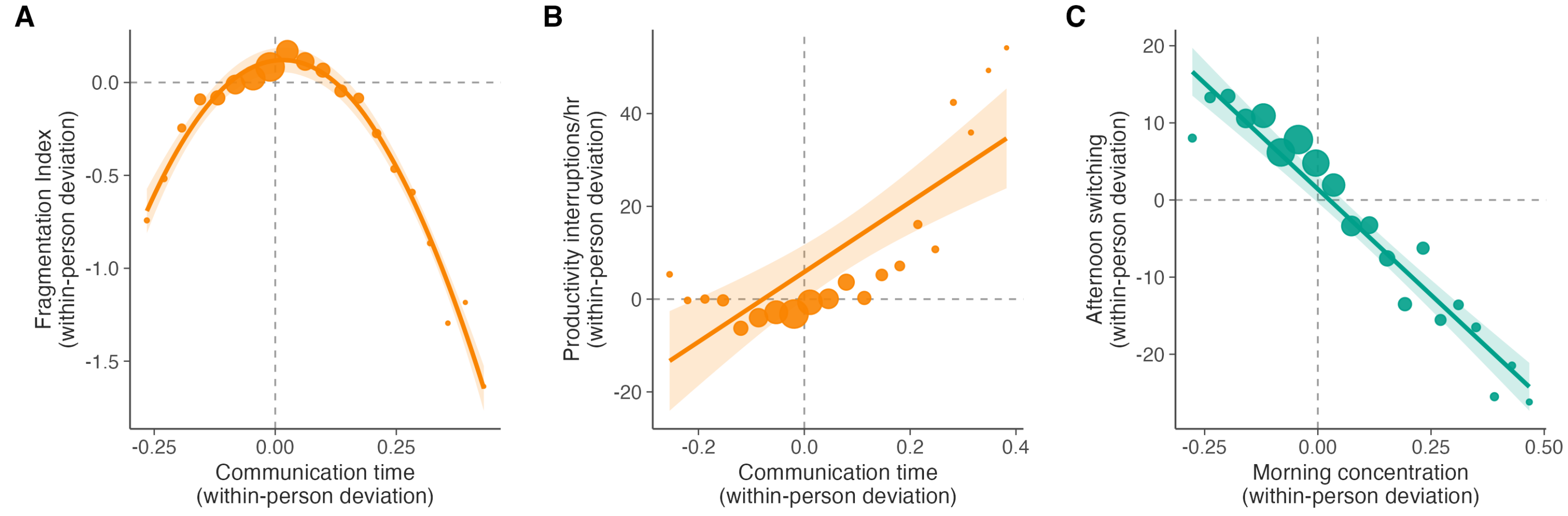

**Fig. 6 | Communication demand and morning concentration coincide with day-to-day shifts in afternoon fragmentation. a.** Within-person association between time spent in communication applications and the Fragmentation Index. The relationship is non-monotonic, with fragmentation rising at moderate levels of communication and declining at high levels as communication comes to dominate the day. Point size is proportional to the number of person-days at each bin. **b.** Within-person association between communication time and the rate of productivity interruptions per hour. More communication-heavy days carry more interruptions of productive work. **c.** Within-person association between morning concentration (the Herfindahl-Hirschman concentration of activity across application categories before noon) and afternoon switching rate. Mornings spent in more concentrated activity predict less fragmented afternoons.

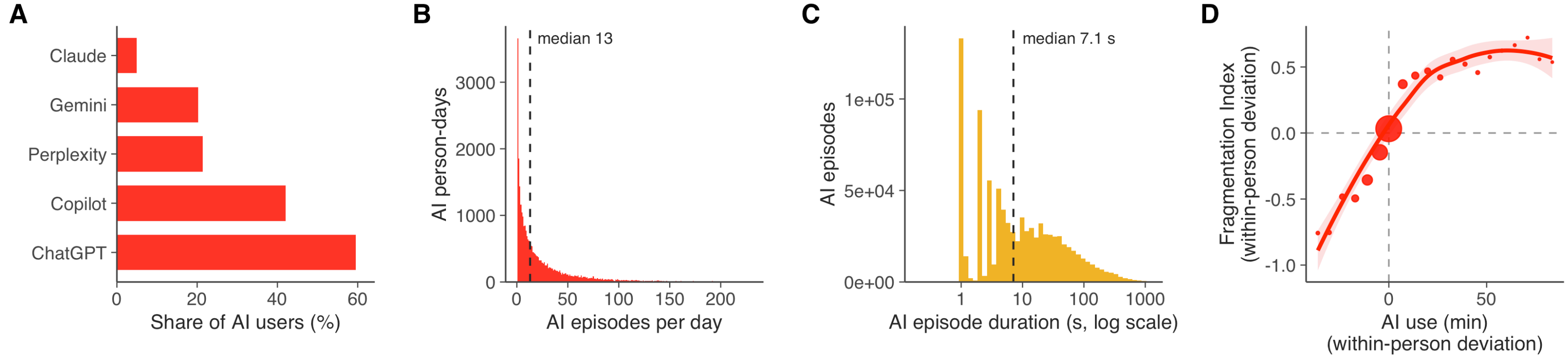

**Fig. 7 | Generative AI tool use is concentrated among a few platforms, occurs in brief and frequent episodes, and coincides with greater within-person daily fragmentation. a.** Share of total AI events attributable to each of five generative AI tools, with the number of distinct users in parentheses. ChatGPT accounts for 70% of AI events; Microsoft Copilot, Google Gemini, Perplexity, and Anthropic Claude account for the remainder. **b.** Distribution of the number of AI episodes per workday across employees who used an AI tool that day. The dashed line marks the median of 13 episodes per day. **c.** Distribution of individual AI episode durations on a log scale, with the dashed line marking the median of 7 seconds. Most episodes are brief, with a long tail of longer sessions. **d.** Within-person association between daily AI use (in minutes) and the Fragmentation Index. Point size is proportional to the number of person-days at each bin. Days on which employees used AI for longer durations are also days on which their digital fragmentation was higher.

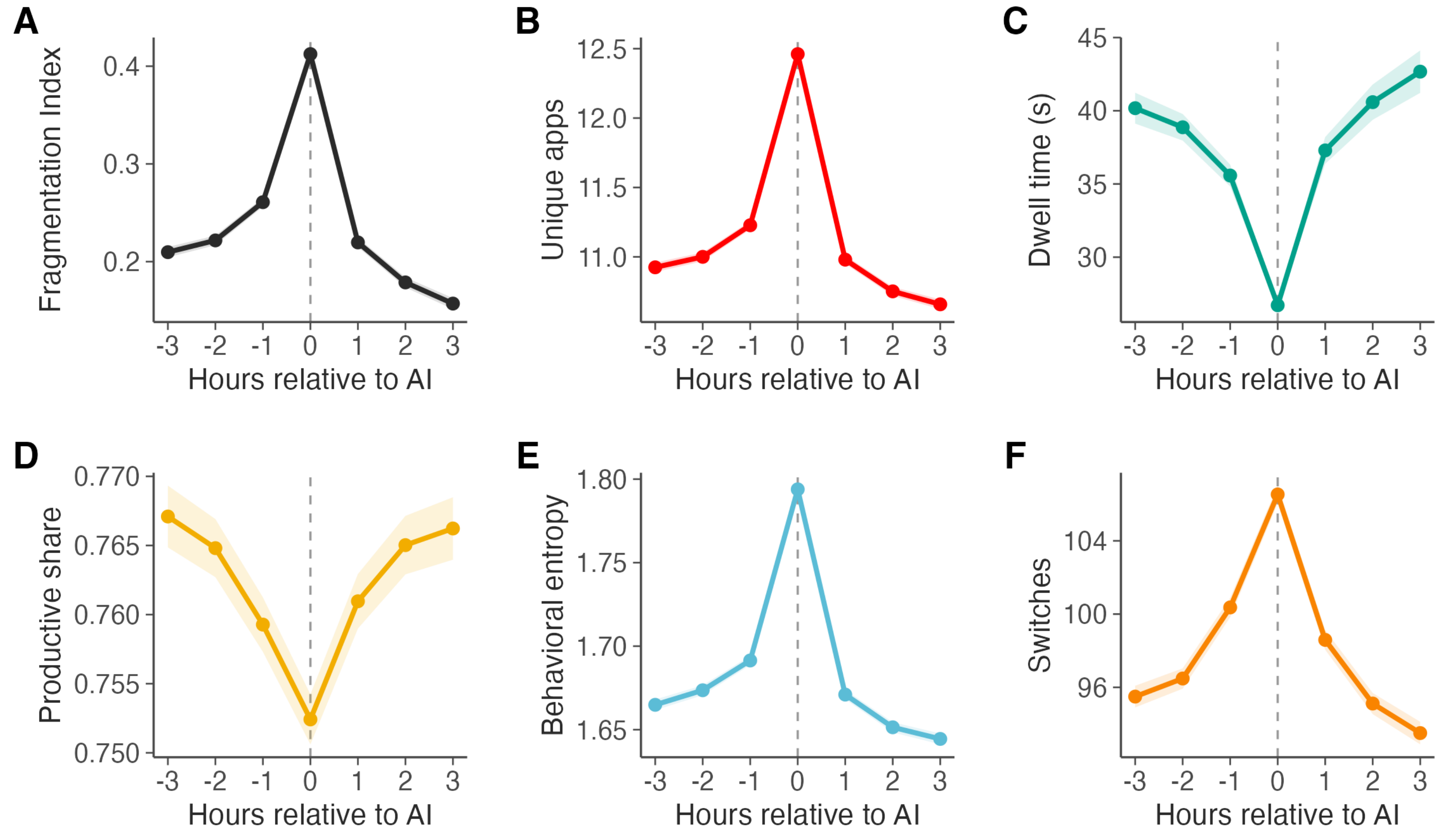

**Fig. 8 |** Generative AI use coincides with the most fragmented hour of the workday, which then subsides. a, Mean Fragmentation Index by hour relative to AI use; hour 0 is the hour of AI use. The Index rises to a peak at the hour of AI use and declines over the following hours. b–f, Components of the workday's switching structure across the same hours — the number of unique applications (b), the application-switch rate (c), the mean time spent in each application (d), the share of time in productive applications (e), and the entropy of application use (f). Each measure peaks at the hour of AI use (dwell time troughs) and returns toward baseline over the

subsequent hours. The ±10-event window contrast around each individual AI use is reported as a robustness analysis (Supplementary Tables S22 and S44).